%% file: main.tex
\documentclass[conference,10pt
]{IEEEtran}
\IEEEoverridecommandlockouts    
\usepackage{verbatim} 
\usepackage{cite}
\usepackage{array}
\usepackage{bm}
\usepackage{amsmath,amssymb,amsfonts}
\usepackage{algorithmic}
\usepackage{graphicx}
\usepackage{tikz}
\usepackage{circuitikz}
\usepackage{siunitx}
\usepackage{gensymb}
\usepackage{makecell}
\usetikzlibrary{shapes.geometric, arrows,chains, positioning}
\usetikzlibrary{arrows.meta} 
\usetikzlibrary { decorations.pathmorphing, decorations.pathreplacing, decorations.shapes, } 
\usepackage{pbox}
\usepackage{relsize}
\tikzset{
    pin/.style = {font = \relsize{-2}} 
}
\ctikzset{
    bipoles/length = 4em, 
    font = \relsize{-1}, 
}
\usepackage{textcomp}
\usepackage{xcolor}
\def\BibTeX{{\rm B\kern-.05em{\sc i\kern-.025em b}\kern-.08em
    T\kern-.1667em\lower.7ex\hbox{E}\kern-.125emX}}
\usepackage{mathtools}
\usepackage{schemabloc}
\usetikzlibrary{circuits}
\usetikzlibrary{arrows}
\makeatletter
\let\MYcaption\@makecaption
\makeatother
\usepackage[font=footnotesize]{subcaption}
\makeatletter
\let\@makecaption\MYcaption
\makeatother
\usepackage{caption}

\usepackage{pgfplotstable}
\usepackage{filecontents} 

\usepackage{multirow}
\usepackage{mathtools}
\usepackage{commath}

\newcommand{\RNum}[1]{\uppercase\expandafter{\romannumeral #1\relax}}
\graphicspath{ {./confimages/} }
 \usepackage{pgfplots}
  \pgfplotsset{compat=newest}
  \usetikzlibrary{plotmarks}
  \usetikzlibrary{arrows.meta}
  \usepgfplotslibrary{patchplots}
  \usepackage{grffile}
  \usepackage{amsmath}
  
\usepackage{amsthm}

\usepackage{fancyhdr,graphicx,amsmath,amssymb}
\usepackage[ruled,vlined]{algorithm2e}
\usepackage{times}
\usepackage{hhline}

\usepackage{hhline}
\usepackage{booktabs} 
\usepackage{circuitikz}
\usepackage{siunitx}
\usepackage[inline]{enumitem}

\usepackage[hyphens]{url}
\usepackage{balance}
\usepackage[commandnameprefix=always]{changes}
\definechangesauthor[color=blue,name=Vassilis]{VP}
\usepackage{nicematrix}

\usepackage{hyperref}
\hypersetup{
    colorlinks=true,
    linkcolor=blue!70,
    filecolor=magenta,      
    urlcolor=cyan,
    citecolor=teal!77,
    pdftitle={RISaided_Positioning_Sensing_Journal}
}



\begin{document}

\allowdisplaybreaks 

\title{Optimizing Indoor RIS-Aided Physical-Layer Security: A Codebook-Generation Methodology and Measurement-Based Analysis\vspace{-0.17cm}}

\author{\IEEEauthorblockN{ 
Dimitris Kompostiotis\IEEEauthorrefmark{2}, Dimitris Vordonis\IEEEauthorrefmark{2}, Vassilis Paliouras\IEEEauthorrefmark{2}, and George C. Alexandropoulos\IEEEauthorrefmark{5}}
\IEEEauthorblockA{\IEEEauthorrefmark{2}Electrical and Computer Engineering Department, University of Patras, Greece,\\
 \IEEEauthorrefmark{5}Department of Informatics and Telecommunications, National and Kapodistrian University of Athens, Greece,\\
e-mails: \{d.kompostiotis, d.vordonis\}@ac.upatras.gr, paliuras@ece.upatras.gr, alexandg@di.uoa.gr}\vspace{-0.77 cm}}

\maketitle
\begin{abstract}
Sixth-Generation (6G) wireless networks aim to support innovative Internet-of-Things (IoT) applications that demand faster and more secure data transmission. While higher Open Systems Interconnection (OSI) layers employ measures like encryption and secure protocols to address data security, Physical-Layer Security (PLS) focuses on preventing information leakage to EavesDroppers (EDs) and mitigating the effects of jammers and spoofing attacks. In this context, the emerging technology of Reconfigurable Intelligent Surfaces (RISs) can play an instrumental role, enhancing PLS by intelligently reflecting electromagnetic waves to benefit Legitimate Users (LUs) while obstructing EDs. This paper presents practical indoor measurements to evaluate the capability of an RIS to enhance PLS, focusing on a varactor-based RIS technology designed for the FR1 band at $3.55$~GHz. A comparative analysis of state-of-the-art RIS-aided secrecy optimization algorithms together with a novel approach designed in this paper, which relies on a newly generated RIS phase configuration codebook, highlight the potential of RISs to improve both data rates for LUs as well as secrecy against EDs in real-world indoor multipath environments. The results also demonstrate the frequency selectivity of the RIS, proviging practical insights on the optimization of the technology.
\vspace{-0.3cm}
\end{abstract}


\section{Introduction}
\label{sec:Intro}
One of the most challenging issues in wireless communication networks is ensuring secure connections between nodes. In the context of the Open Systems Interconnection (OSI) model for current networks and for their upcoming sixth Generation (6G), each layer above the physical layer addresses specific aspects of data security. These include maintaining confidentiality, integrity, authentication, and non-repudiation of data as it traverses the network~\cite{stapleton2014security}. To mitigate threats unique to each layer, a variety of security measures is employed, such as encryption, firewalls, secure protocols, etc.~\cite{stewart2013network}.

While security concerns are extensively addressed and largely resolved by higher OSI layers in current wireless networks, Physical-Layer Security (PLS) remains a significant challenge. Beyond safeguarding the hardware and transmission media against unauthorized access and tampering, PLS is also focused on ensuring secure and reliable communications by preventing information leakage to potential EavesDroppers (EDs)~\cite{zhou2013physical} and mitigating the impact of jammers~\cite{zhou2013physical}. This domain of security can be categorized into two main areas~\cite{7136233}: managing passive eavesdropping~\cite{kompostiotis2023secrecy,kompostiotis2024secrecy}, where an unauthorized user (or ED) attempts to intercept signals or information exchanged between legitimate entities; and addressing jamming (or active eavesdropping), where an active adversary (or jammer) transmits corrupted signals to disrupt communication with Legitimate Users (LUs)~\cite{7136233}. A sub-case of jamming attack is the so-called spoofing~\cite{yilmaz2015survey}, where the spoofer starts transmitting a deceiving signal to the LU by pretending to be the legitimate Transmitter (Tx).

The main objective of PLS is to maximize the rate of reliable information from the source to the intended destination, while all malicious nodes are kept as ignorant of this information as possible~\cite{wyner1975wire,bloch2008wireless}. This maximum reliable rate is known as Secrecy Capacity (SC)~\cite{wyner1975wire,bloch2008wireless}. Wyner~\cite{wyner1975wire} defined the wiretap channel and established the possibility to create almost perfect secure communication links without relying on private keys. Exploiting the Additive White Gaussian Noise (AWGN) and stochastic encoders (which map each message to many codewords), maximal uncertainty is induced in the ED~\cite{wyner1975wire}. A follow-up work~\cite{leung1978gaussian} characterizes the SC of the AWGN wiretap channel. After that, considerable efforts have been made to generalize these studies to the wireless channel, also considering multi-user scenarios~\cite{khisti2007gaussian,oggier2011secrecy,shafiee2007achievable}. The most challenging part in optimizing SC is that the passive ED's Channel State Information (CSI) is unknown, which is not the case for the active ED, since an active ED transmits signals to the network, and thus its existence can be estimated~\cite{al2016physical}. This is why many works on passive ED attacks assume either statistical knowledge of the ED's CSI~\cite{pei2011secure,10143983} or that the ED's CSI is known, only when the ED is an active user in the system, but untrusted from the LU~\cite{mukherjee2014principles,AKW2021}.

Today, advances in microelectromechanical systems and metamaterials have introduced innovative hardware components for the physical layer of wireless communications~\cite{WavePropTCCN}. A promising one is the technology of Reconfigurable Intelligent Surfaces (RISs)~\cite{huang2019reconfigurable}, which constitutes of planar arrays made up of metamaterials with adjustable properties. RISs can dynamically control the reflection of incoming electromagnetic waves in the desired way~\cite{RIS_challenges_all,kompostiotis2023received,vordonis2022reconfigurable} and mainly rely on nearly passive metamaterials, that do not require power amplifiers or radio-frequency chains, making them a cost-effective and energy-efficient solution for 6G wireless networks~\cite{RISE6G_COMMAG}.

In recent years, RIS-aided PLS has gained considerable research interest and significant growth~\cite{wu2021intelligent,kompostiotis2023secrecy,cui2019secure,yu2020robust,zhou2021intelligent,guan2020intelligent,jiang2021joint,AKW2021,10143983,RIS_occultation} relying mainly on simulations. For example, RIS elements have been optimized to absorb jamming signals~\cite{10296481}, reflect artificial noise~\cite{guan2020intelligent,10143983,arzykulov2023artificial}, and disrupt the reflection of information signals towards potential EDs~\cite{cui2019secure}. An important challenge in designing RISs for PLS is to simultaneously maximize the data rate for the LU while minimizing information leakage to the ED, especially when the channels are correlated~\cite{secrecy_corr}. However, only a few works have relied on actual measurements to showcase the feasibility and potential of RIS-aided PLS. For example,~\cite{kayraklik2024indoor} presented an experimental setup demonstrating the RIS capabilities to enhance signal coverage and provide PLS in an indoor environment, using a PIN-diode-based dual-polarized RIS at $5.2$~GHz. 

In this paper, we present an indoor RIS-aided proof of concept at $3.55$~GHz for RIS-aided PLS. Three are the main contributions of this work: \begin{enumerate*}[label={\textbf{(\alph*)}}]
\item unlike~\cite{kayraklik2024indoor}, we herein experiment with a varactor-based dual-polarized RIS designed for the FR1 band and present comparative results between two basic algorithms that optimize SC. The first utilizes the entire RIS to enhance the LU’s signal, while weakening the ED’s signal. In contrast, the second algorithm partitions the RIS into two regions, each dedicated to serving the intended and unintended users separately, as in~\cite{kayraklik2024indoor}. Furthermore, two more methods are included for baseline comparison: one aiming to optimize the LU's channel and the other aiming at degrading the ED's channel, ignoring the channels of the ED and the LU, respectively. \item Additionally, we introduce a novel methodology for designing an RIS phase configuration codebook for indoor multipath environments, where a specific RIS configuration is selected each time an LU communicates with the Access Point (AP). Each configuration in the codebook is designed not only to increase the received power--and consequently the data rate--for the LU, but also to ensure secrecy against potential EDs. \item Finally, insights about the issues occurring from the frequency selectivity of the RIS, are sketched. \end{enumerate*} 

The remainder of the paper is organized as follows: Section~\ref{s:systemmodel} describes the signal model and setup used for the RIS-aided proof-of-concept, while Sections~\ref{s:codebooks_gen} and~\ref{s:measurements} introduce the RIS phase configuration codebook generation and the field test results, respectively. 
Finally, Section~\ref{s:conclusion} concludes the paper.

\section{Indoor RIS-Aided PLS Setup}
\label{s:systemmodel}
The considered RIS-aided PLS setup is depicted in Fig.~\ref{fig:Channel_Model}. It consists of a single-antenna AP-Tx which communicates via Orthogonal Frequency Division Multiplexing (OFDM) with a single-antenna LU in the presence of an ED. An RIS, with $M{\triangleq} N_V{\times}N_H$ metamaterials ($m{=}1,{\ldots},M$ is the element index), is deployed to enable wireless communications between the AP and the LU; however, it also establishes a link between the AP and the ED. The considered RIS is also equipped with a controller~\cite{RIS_challenges_all} that coordinates the AP and the metasurface for channel acquisition and data transmission~\cite{wu2021intelligent}. 

All channels address multipath frequency-selective fading. We denote by $h^{p}_{d}[v]{\in}\mathbb{C}$ the direct channel from the AP to LU/ED ($p{=}l$ for LU and $p{=}e$ for ED) for a specific subcarrier index $v$, $\bm{h}^{p}_{\text{RIS}}[v]{\in} \mathbb{C}^{1 {\times} M}$ is the RIS-to-LU/ED channel matrix, and $\bm{g}_{\text{RIS}}[v]{\in} \mathbb{C}^{M {\times} 1}$ is the AP-to-RIS channel matrix. All aforementioned channel matrices are the discrete complex baseband representations around a specific carrier frequency; in our case the FR1 frequency of $3.55$~GHz. After transmitting the discrete-frequency baseband signal $x[v]{\in}\mathbb{C}$, the received signals at the LU, $y^{l}[v] {\in} \mathbb{C}$, and ED, $y^{e}[v] {\in} \mathbb{C}$, are given by:
\begin{figure}[t]
    \centering
    \scalebox{0.14}{\includegraphics{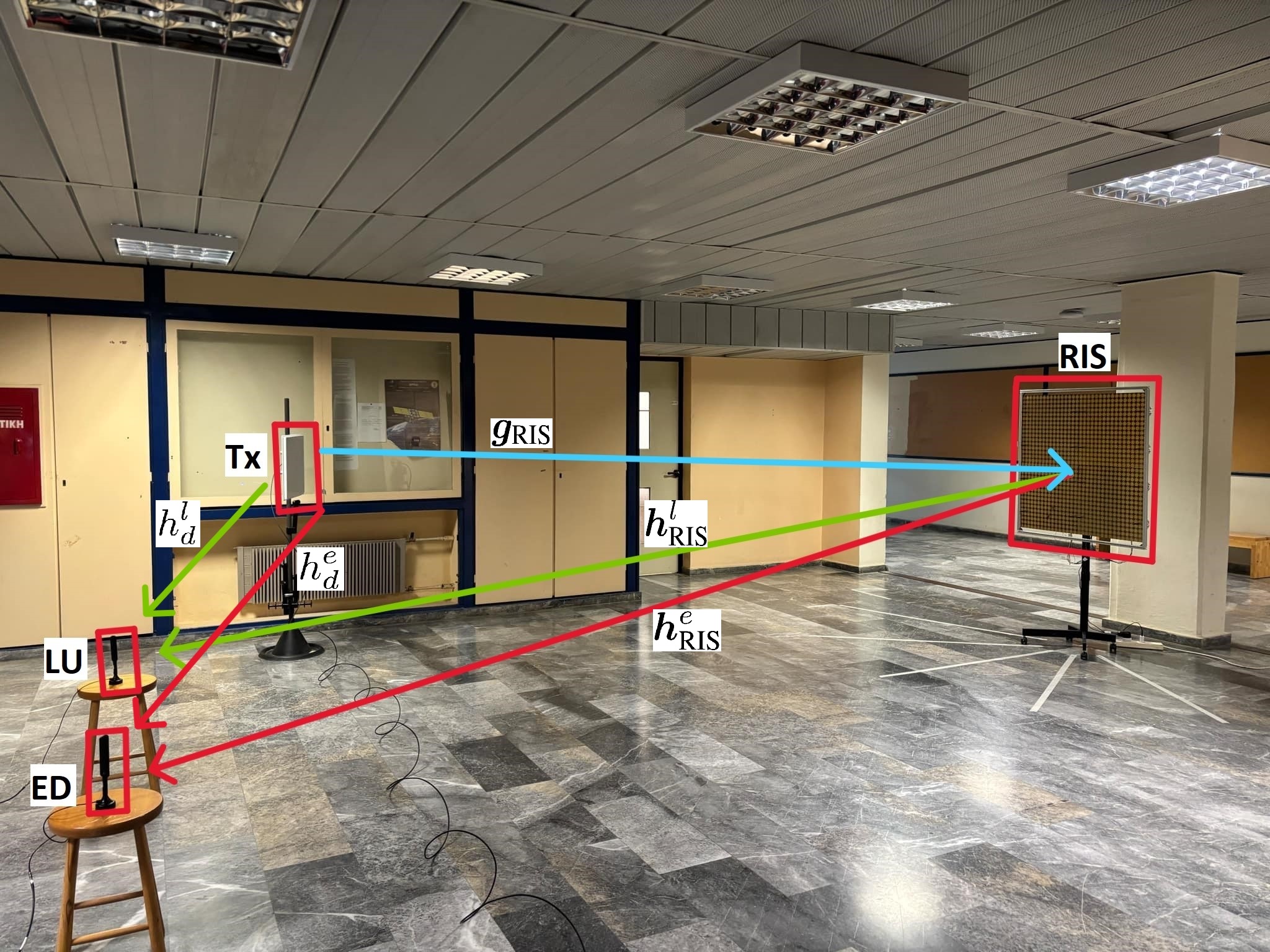}}
    \caption{Measurement setup of the designed RIS-enabled PLS system.}
    \label{fig:Channel_Model}
\end{figure}
\begin{equation}
\begin{aligned}
    y^{l}[v] &= \left(h^{l}_{d}[v] + \bm{h}^{l}_{\text{RIS}}[v]\bm{\Phi_c}[v]\bm{g}_{\text{RIS}}[v]\right)x[v] + n^{l}[v]\\
    y^{e}[v] &= \left(h^{e}_{d}[v] + \bm{h}^{e}_{\text{RIS}}[v]\bm{\Phi_c}[v]\bm{g}_{\text{RIS}}[v]\right)x[v] + n^{e}[v],
\end{aligned}
\label{eq:received_signals}
\end{equation}
where, for each subcarrier $v$, $n^{p}[v]$ denotes the zero-mean AWGN with variance $N_0$ (assumed the same for LU and ED), 
and $\bm{\Phi_c}[v]$ is the RIS configuration response, defined as:
\begin{align*}
 \bm{\Phi_c}[v] {\triangleq} \operatorname{diag}\left\{\alpha_{1}\!\left(v,c_1\right)e^{\jmath\theta_{1}\left(v,c_1\right)}, \ldots ,\alpha_{M}\!\left(v,c_M\right)e^{\jmath\theta_{M}\left(v,c_M\right)}\right\}
\end{align*}
when the configuration $\bm{c}{\triangleq}[c_1,\ldots,c_M]^{T}$, with $c_m{\in}\{0,1\}$~$\forall m$, is applied on the metasurface. Since $c_m$ can assume two possible values and it holds $\alpha_{m}\left(v,c_m\right){\approx} 1,$~$\forall(m,v)$ from the respective measurements in~\cite{rains2023fully}, each metamaterial can add binary phase shifts to the incoming signal. In addition, for the phase response of each $m$-th element, it holds $\theta_m\left(v,c_m\right) {\in} [0,\pi]$~$\forall (m,v)$.
The latter implies that the overall RIS configuration is performed per the whole bandwidth of influence of the metasurface~\cite{RIS_challenges_all}, and not per subcarrier. 
It is also worth noting that, in our measurement setup illustrated in Fig.~\ref{fig:Channel_Model}, the direct channel component $h^{p}_{d}[v]$ of the received signal was power-limited, since a directive antenna was used to transmit signals towards the RIS. Therefore, the direct link between the Tx and LU/ED terminals was out of the antennas' beamwidth. The overall list of components used in our measurement setup for transmission and reception, with their corresponding descriptions, is given in Table~\ref{tab:hardware_components}.

\begin{table}[tb]
\caption{Hardware specifications of the experimental setup in Fig.~\ref{fig:Channel_Model}.\vspace*{-0.5cm}}
\begin{center}
\scalebox{1.0}
{
\setlength{\tabcolsep}{2pt}
\begin{tabular}{p{0.17\linewidth}  p{0.77\linewidth}}
\toprule
\toprule
\textbf{Hardware}& \textbf{Description} \\
\midrule
\midrule
USRP X300 & This Software-Defined Radio (SDR) platform was used for signal transmission at the carrier frequency of $3.55$~GHz with power amplification gain $20$~dB. \\
\midrule
Flat-panel PAT3519XP antenna & This ITELITE model comprises dual-polarized antennas, featuring a beamwidth of approximately $20^\circ$ at the frequency range $[3.5,3.8]$~GHz while minimizing the effects of unwanted reflections and interference.\\
\midrule
RIS~\cite{rains2023fully} & The metasurface comprises four tiles, each featuring a $16 {\times} 16$ array of unit cells. Each unit cell is equipped with two independently addressable varactor diodes, controlled by programmable bias voltages set at $11$ and $7.5$~Volts. 
\\
\midrule
USRP N310 & This SDR platform was deployed for signal reception at the carrier frequency of $3.55$~GHz with sampling frequency at $1$~MHz, offering a gain of $45$~dB. \\ 
\midrule
\makecell[tl]{Omni\\antennas} & Both the LU and the ED used common omnidirectional antennas for signal reception.\\
\bottomrule
\bottomrule
\end{tabular}
}
\label{tab:hardware_components}
\end{center}
\end{table}

\subsection{RIS Design Objective}
Our objective for the RIS in this paper is to design its configuration $\bm{c}$, that leads to a RIS-configuration frequency response $\bm{\Phi_c}[v]$ for each $v$-th subcarrier (let $K$ denote the total number of subcarriers per OFDM symbol, thus, $v=1,\ldots,K$) with the goal to maximize the achievable Secrecy Spectral Efficiency (SSE)~\cite{kompostiotis2024secrecy,AKW2021,10143983}. We have particularly focused on the following sum-SSE metric: 
\begin{align*}
 \kern-0ptR_{\text{sec}} & \triangleq \left[ \sum_{v{=}1}^{K} \left(R_l[v]{-} R_e[v] \right) \, \right]^{+}  \nonumber
 \\&{=} \left[ \sum_{v{=}1}^{K}\!\! \left(\!\log_2\left( 1 {+} \dfrac{\left\Vert(h^{l}_{d}[v] {+} \bm{h}^{l}_{\text{RIS}}[v]\bm{\Phi_c}[v]\bm{g}_{\text{RIS}}[v])x[v]\right\Vert^{2}}{N_0}\right)\right.\right. \notag\\ &
 \quad\left.\left.{-} \log_2\left(1 {+}\dfrac{\left\Vert(h^{e}_{d}[v] {+} \bm{h}^{e}_{\text{RIS}}[v]\bm{\Phi_c}[v]\bm{g}_{\text{RIS}}[v])x[v]\right\Vert^{2}}{N_0}\right)\right)\!\!\right]^{\smash{\mathrlap{+}}}\!,
\end{align*}
where $R_l[v]$ and $R_e[v]$ denote the per sub-carrier spectral efficiencies for the legitimate and eavesdropping links, respectively, and $[x]^{+}{\triangleq}\operatorname{max}\{0,x\}$. Clearly, a higher SSE can be obtained when increasing the power of the LU's received signal while decreasing the power that reaches ED. \vspace{0.05cm}
\subsubsection{Narrowband Transmissions}
\vspace{-0.05cm}
Considering an RIS implementation, with a uniform frequency response across all subcarriers of the Tx signal, different SSE-optimization algorithms can be fairly compared, allowing the selection of the best one in terms of both performance and computational complexity. However, since the actual RIS~\cite{rains2023fully} used here exhibits a frequency-selective response, to draw relevant conclusions about the various algorithms, a sinusoidal (narrowband) signal was initially used as the Tx signal. In this case, the signal $x$ in \eqref{eq:received_signals} was a complex sine wave of a frequency of $100$~KHz (corresponding to a single subcarrier index $v$), and both the corresponding wireless channels and the RIS response were considered as narrowband. 
\subsubsection{OFDM Transmissions}
\vspace{-0.07cm}
On the other hand, to examine the frequency selectivity of the designed RIS-aided PLS setup, the wideband Positioning Reference Signal (PRS) described in the late 3rd generation partnership project's (3GPP) standards was used~\cite{shamaei2021receiver}. Specifically, 5G downlink transmissions uses OFDM with a Cyclic Prefix (CP). Each $10$~ms frame is divided into $10$ subframes of $1$~ms each, or alternatively into two half-frames (subframes $0–4$ and $5–9$), allowing coexistence with LTE systems. Each subframe contains multiple slots, and with normal CP, each slot includes $14$ OFDM symbols. The number of slots per subframe depends on the subcarrier spacing, which is flexible in 5G due to the defined numerologies $\mu{\in}\{0,1,2,3,4\}$, leading to the subcarrier spacings $\Delta_{f}{=}2^{\mu}15$~KHz. Consequently, each subframe contains $2^{\mu}$ slots, and the CP length is proportionally shortened compared to LTE~\cite{shamaei2021receiver}. Subcarrier spacings of $15$ or $30$ KHz are best suited for FR1 (lower frequencies and larger cells), while higher spacings are used for FR2 (higher frequencies). Finally, in the frequency domain, each subframe consists of resource grids, which are further divided into resource blocks (each with 12 subcarriers). The smallest unit is a resource element defined by a subcarrier and symbol index.

In this paper, to evaluate the frequency selectivity of the deployed RIS device, we have adopted the middle numerology $\mu{=}2$, resulting in a subcarrier spacing equal to $60$~KHz, adding also extended CP, which means that $12$ OFDM symbols were included in a single slot. In the frequency dimension of the Tx signal, $52$ resource blocks were used, where each one contained $12$ subcarriers per OFDM symbol. However, here, only $\frac{12}{\mu}{=}\frac{12}{2}{=}6$ were occupied due to the selected numerology. The resulting bandwidth of the Tx signal was thus $B{=}52{\times}12{\times}60{\times}10^{3}{=}37.44$~MHz, since we used $52$ resource blocks, $12$  subcarriers per resource block, and a subcarrier spacing of $60$~KHz.

\section{Proposed Methodology for Codebook Design}
\label{s:codebooks_gen}
Aiming at computing the RIS configuration $\bm{c}$ that maximizes SSE, we first introduce Algorithm~\ref{algo:1} that maximizes the ratio between the LU's and ED's received powers, $P_r$, using the entire RIS panel. The algorithm traverses the columns and rows of the RIS, changing each column's and row's reflection phase separately. It then computes the achievable $P_r$ and checks whether each new configuration achieves a ratio value greater than the previous one. This process is continued till scanning all RIS columns and rows to identify the $\bm{c}$ configuration maximizing $P_r$.
\begin{algorithm}[t]
\footnotesize
\KwIn{Initialize the RIS configuration as $\bm{c}=\bm{0}_{N_H\times N_V}$.}
\KwOut{The RIS configuration $\bm{c}$ maximizing the ratio between the LU's and ED's received powers.}
\nl Capture data for both the LU and ED using the USRP N310\;
\nl Measure the LU's and ED's received powers: 
$\Vert(h^{p}_{d} {+} \bm{h}^{p}_{\text{RIS}}\bm{\Phi_c}\bm{g}_{\text{RIS}})x\Vert^{2}$ for $p{=}l$ and $p{=}e$, respectively\;
\nl $P_{r}^{\rm old} = \Vert(h^{l}_{d} {+} \bm{h}^{l}_{\text{RIS}}\bm{\Phi_c}\bm{g}_{\text{RIS}})x\Vert^{2} / \Vert(h^{e}_{d} {+} \bm{h}^{e}_{\text{RIS}}\bm{\Phi_c}\bm{g}_{\text{RIS}})x\Vert^{2}$\;
\nl \For {\texttt{$iter=1,2$}}{
        \nl \For {\texttt{$col{=}1,\ldots, N_H$}}{
        \nl Reverse RIS's column $col$ state\;
        \nl Capture data for both the LU and ED\;
        \nl Measure the LU's and ED's received powers:
        
        $\Vert(h^{p}_{d} {+} \bm{h}^{p}_{\text{RIS}}\bm{\Phi_c}\bm{g}_{\text{RIS}})x\Vert^{2}$ for $p{=}l$ and $p{=}e$, respectively\;
        \nl $P_{r}^{\rm new} {=} \Vert(h^{l}_{d} {+} \bm{h}^{l}_{\text{RIS}}\bm{\Phi_c}\bm{g}_{\text{RIS}})x\Vert^{2} / \Vert(h^{e}_{d} {+} \bm{h}^{e}_{\text{RIS}}\bm{\Phi_c}\bm{g}_{\text{RIS}})x\Vert^{2}$\;
        \nl \If{$P_{r}^{\rm new} \leq P_{r}^{\rm old}$}{ 
            \nl Reverse again RIS's column $col$ state\;
        }
        \nl \lIf{$P_{r}^{\rm new} > P_{r}^{\rm old}$}{
             $P_{r}^{\rm old} = P_{r}^{\rm new}$
        }
    }
    \nl \For {\texttt{$row{=}1,\ldots, N_V$}}{
        \nl Reverse RIS's row $row$ state\;
        \nl Capture data for both the LU and ED\;
        \nl Measure the LU's and ED's received powers: 
        
        $\Vert(h^{p}_{d} {+} \bm{h}^{p}_{\text{RIS}}\bm{\Phi_c}\bm{g}_{\text{RIS}})x\Vert^{2}$ for $p{=}l$ and $p{=}e$, respectively\;
        \nl $P_{r}^{\rm new} {=} \Vert(h^{l}_{d} {+} \bm{h}^{l}_{\text{RIS}}\bm{\Phi_c}\bm{g}_{\text{RIS}})x\Vert^{2} / \Vert(h^{e}_{d} {+} \bm{h}^{e}_{\text{RIS}}\bm{\Phi_c}\bm{g}_{\text{RIS}})x\Vert^{2}$\;
        \nl \If{$P_{r}^{\rm new} \leq P_{r}^{\rm old}$}{ 
            \nl Reverse again RIS's row $row$ state\;
        }
        \nl \lIf{$P_{r}^{\rm new} > P_{r}^{\rm old}$}{
            $P_{r}^{\rm old} = P_{r}^{\rm new}$
        }
    }
}
\caption{SSE Maximization via the Entire RIS}\label{algo:1}
\end{algorithm}
On the other hand, Algorithm~\ref{algo:2}~\cite{kayraklik2024indoor} partitions the RIS into two regions and optimizes each part separately. One RIS region is devoted to the LU to maximize its received power, while the other minimizes the ED's received power, respectively. It is worth noting that, power computation in the wideband case follows Algorithms~\ref{algo:1} and~\ref{algo:2} on a per-subcarrier basis and is then aggregated over the entire bandwidth. Finally, Algorithms~\ref{algo:1} and~\ref{algo:2} will be compared with two additional techniques that perform optimization of the achievable rate. One aiming at maximizing the LU's rate (``LU max.''), by performing beamforming toward the LU ignoring the presence of the ED, while the other seeks to minimize ED's rate (``ED min.''), disregarding the LU. 
\begin{algorithm}[t]
\footnotesize
\KwIn{Initialize the RIS configuration as $\bm{c}=\bm{0}_{N_H\times N_V}$.}
\KwOut{The RIS configuration $\bm{c}$ maximizing the ratio between the LU's and ED's received powers.}
\nl Capture data for both the LU and ED using the USRP N310\;
\nl Measure the LU's and ED's received powers: 

$\Vert(h^{p}_{d} {+} \bm{h}^{p}_{\text{RIS}}\bm{\Phi_c}\bm{g}_{\text{RIS}})x\Vert^{2}$ for $p{=}l$ and $p{=}e$, respectively\;
\nl $P_{\rm LU}^{\rm old} {=} \Vert(h^{l}_{d} {+} \bm{h}^{l}_{\text{RIS}}\bm{\Phi_c}\bm{g}_{\text{RIS}})x\Vert^{2}$, 
$P_{\rm ED}^{\rm old} {=} \Vert(h^{e}_{d} {+} \bm{h}^{e}_{\text{RIS}}\bm{\Phi_c}\bm{g}_{\text{RIS}})x\Vert^{2}$\;
\nl \For {\texttt{$iter=1,2$}}{
        \nl \For {\texttt{$col{=}1,\ldots, \frac{N_H}{2}$}}{
        \nl Reverse RIS's column $col$ state\;
        \nl Capture and measure $P_{\rm LU}^{\rm new} = \Vert(h^{l}_{d} {+} \bm{h}^{l}_{\text{RIS}}\bm{\Phi_c}\bm{g}_{\text{RIS}})x\Vert^{2}$\;
        \nl \If{$P_{\rm LU}^{\rm new} \leq P_{\rm LU}^{\rm old}$}{ 
            \nl Reverse again RIS's column $col$ state\;
        }
        \nl \lIf{$P_{\rm LU}^{\rm new} > P_{\rm LU}^{\rm old}$}{
             $P_{\rm LU}^{\rm old} = P_{\rm LU}^{\rm new}$
        }
    }
    \nl \For {\texttt{$col{=}\frac{N_H}{2}+1,\ldots, N_H$}}{
        \nl Reverse RIS's column $col$ state\;
        \nl Capture and measure: $P_{\rm ED}^{\rm new} = \Vert(h^{e}_{d} {+} \bm{h}^{e}_{\text{RIS}}\bm{\Phi_c}\bm{g}_{\text{RIS}})x\Vert^{2}$\;
        \nl \If{$P_{\rm ED}^{\rm new} \geq P_{\rm ED}^{\rm old}$}{ 
            \nl Reverse again RIS's column $col$ state\;
        }
        \nl \lIf{$P_{\rm ED}^{\rm new} < P_{\rm ED}^{\rm old}$}{
             $P_{\rm ED}^{\rm old} = P_{\rm ED}^{\rm new}$
        }
    }
    \nl \For {\texttt{$row{=}1,\ldots, N_V$}}{
        \nl Reverse half RIS's (LU's) row $row$ state\;
        \nl Capture and measure $P_{\rm LU}^{\rm new} = \Vert(h^{l}_{d} {+} \bm{h}^{l}_{\text{RIS}}\bm{\Phi_c}\bm{g}_{\text{RIS}})x\Vert^{2}$\;
        \nl \If{$P_{\rm LU}^{\rm new} \leq P_{\rm LU}^{\rm old}$}{ 
            \nl Retrieve the half RIS's (LU's) row $row$ state\;
        }
        \nl \lIf{$P_{\rm LU}^{\rm new} > P_{\rm LU}^{\rm old}$}{
             $P_{\rm LU}^{\rm old} = P_{\rm LU}^{\rm new}$
        }
        \nl Reverse other half RIS's (ED's) row $row$ state\;
        \nl Capture and measure $P_{\rm ED}^{\rm new} = \Vert(h^{e}_{d} {+} \bm{h}^{e}_{\text{RIS}}\bm{\Phi_c}\bm{g}_{\text{RIS}})x\Vert^{2}$\;
        \nl \If{$P_{\rm ED}^{\rm new} \geq P_{\rm ED}^{\rm old}$}{ 
            \nl Retrieve the half RIS's (ED's) row $row$ state\;
        }
        \nl \lIf{$P_{\rm ED}^{\rm new} < P_{\rm ED}^{\rm old}$}{
             $P_{\rm ED}^{\rm old} = P_{\rm ED}^{\rm new}$
        }
    }
}
\caption{SSE Maximization via RIS Partition}\label{algo:2}
\end{algorithm}

All the aforementioned approaches obtain a suboptimal solution for the SSE maximization problem for a given LU and ED placement in space. To treat the general case where each of these nodes lies in any point within the RIS indoor area of influence, we have designed a codebook of RIS configurations. To this end, the indoor area shown in Fig.~\ref{fig:Channel_Model} was divided into circular sectors with a $15\degree$ epicenter angle each (see Fig.~\ref{fig:Channel_Model2}), where the center of the circle was the RIS center. The LU and ED were placed in all possible sectors, thus in distinct angles with respect to the RIS. In particular, as shown in Fig.~\ref{fig:Channel_Model2}, the LU (green dots) and ED (red dots) were positioned at a distance of about $7$~m from the RIS center. For each pair of LU, ED positions, we have run all approaches and stored the output RIS configurations in the codebook $\mathcal{CB}$. This implies that, for any relative LU position with respect to the RIS, there exists a good RIS configuration independently of the ED's location.

The question that still remains is how to decide which RIS configuration in $\mathcal{CB}$ is appropriate for achieving better SSE each time. Depending on the actual positions of the LU and ED (when, e.g., the ED is an active user in the system), the RIS configuration computed for the closest corresponding locations can be used each time. However, this is particularly challenging, especially for RIS implementations with~\mbox{$1$-bit} quantization of element reflection phase, since side-lobe effects take place~\cite{vordonis2025evaluating,RIS_beams}. Specifically, although the algorithms are capable to achieve a good power separation between the LU and ED (thus, non-zero SSE), this is only true for certain positions/angles of them. Thus, if one measures the power of the signal reaching different locations in space (assuming that the RIS is configured with a $\bm{c}$ configuration from $\mathcal{CB}$), it can vary significantly, and even exceed, the power of the targeted LU, due to the side-lobe effects. This can be only resolved by measuring with high angle resolution the power pattern of each computed RIS configuration in $\mathcal{CB}$, or by using RIS implementations with higher phase resolution~\cite{RIS_beams}. Finally, even in scenarios where the full CSI of the ED is not available, the proposed method can still be employed to exclude potential critical regions where the ED may be located.

\section{Experimental Results and Discussion}
\label{s:measurements}
\begin{figure}[t]
\vspace{-0.17cm}
    \centering
    \scalebox{1}{\input{confimages/indoor_setting}}
    \vspace{-0.27cm}
    \caption{Top view of the considered indoor setup for RIS-aided PLS. The cyclic sector divided space includes all possible LU and ED locations.\vspace{-0.1cm}}
    \label{fig:Channel_Model2}
\end{figure}
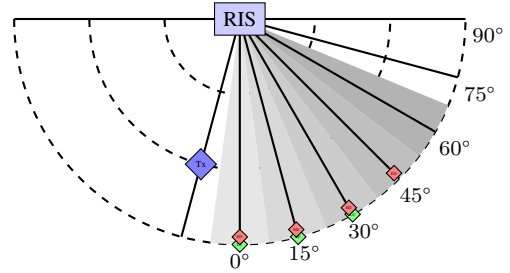
We have obtained field measurements with the PLS system setup illustrated in Fig.~\ref{fig:Channel_Model2}. The experiments were conducted indoors using two SDRs for over-the-air signal transmission and reception and the RIS prototype described in Section~\ref{s:systemmodel}. Two single-antenna receivers were considered, one for the LU and the other for the ED. The Tx's flat-panel antenna was oriented towards the RIS, which created generalized reflections to illuminate the LU and ED antennas. The measurement process comprised two cases: \textit{i}) transmission of a $100$~KHz single-tone sinusoidal signal at the center frequency of $3.55$~GHz, to evaluate the various considered approaches; and \textit{ii}) transmission of the PRS signal described in Section~\ref{s:systemmodel}, with the goal to evaluate the frequency selectivity issues introduced by the RIS panel. The empty room where the experiments took place was divided into a grid according to the angles and distances from the RIS, as shown in Fig.~\ref{fig:Channel_Model2}. The Tx SDR antenna was placed at the fixed location of $-15 \degree$ and $5$~m  from the RIS center, while the LU and ED antennas, representing the two respective users, were placed according to the green and red points, respectively, as shown in Fig.~\ref{fig:Channel_Model2}.
\input{plot_LU-ED_Rec_power_wro_loc}
\input{heatmap}

The SSE performance was evaluated using the $R_{\text{sec}}$ expression, under the assumption that both the LU and ED have identical levels of noise power. 
This assumption has been also considered for the indoor scenarios in the measurement campaigns in~\cite{kecsir2023measurement,kayraklik2024indoor}. We have particularly calculated $R_{\text{sec}}$ without applying the max operator. This decision was made because, in scenarios where the RIS is inactive, the ED's received signal power may exceed that of the LU’s. By ignoring the max operator, the improvement in performance due to the RIS can still be demonstrated even when the unintended user is significantly close to the Tx, as happened in our indoor setup (see Fig.~\ref{fig:Channel_Model}). 
Table~\ref{tab:merged_power_full_grouped} includes the SSE results for all algorithms used. The first part of the table contains the SSE-based optimization approaches, while the second part contains the rate-based ones. In particular, for each possible pair of LU and ED positions, the received powers of the LU and ED have been measured for the following six cases for the RIS configuration: \begin{enumerate*}[label={(\alph*)}] \item when computed via Algorithm~\ref{algo:1} transmitting a sine signal; \item when computed via Algorithm~\ref{algo:1}, yet transmitting the PRS signal; \item when computed via Algorithm~\ref{algo:2} transmitting a sine signal; \item when computed to maximize the LU's received power; \item when computed to minimize the ED's received power; and for \item the case of a uniform RIS surface~\end{enumerate*}. It can be seen from the Table~\ref{tab:merged_power_full_grouped} that the uniform RIS configuration not only did not achieve a noticeable differentiation between the received powers at the LU and ED (in the sense that the power of LU was much larger than that of ED), but in many positions (e.g., LU at $0\degree$ and ED at $15\degree$ or $30\degree$ etc.), the power received by the ED was larger than that of the LU. In contrast, it is shown that both Algorithms~\ref{algo:1} and~\ref{algo:2} achieve an increase in LU's power which happens, for all positions, greater than that of ED's. Moreover, it can be observed that Algorithm~\ref{algo:1} performed a little better than Algorithm~\ref{algo:2} in many cases. The reason is that the second algorithm uses half of the RIS to maximize the LU power, ignoring the effect this configuration has on ED. Similarly, the other half of RIS was used to minimize the ED's received power, disregarding the effect of this half-RIS configuration on the power of the LU's received power. This is the reason why the first algorithm achieves better power separation between the LU and ED, and therefore better SSE performance; this can be also seen in~Fig.~\ref{fig:combined_sec_cap}. Finally, in the case of the PRS transmission, although the LU received higher power than the ED, the power separation between them was not sufficiently strong. This is likely due to the frequency-selective response of the RIS~\cite{vordonis2025evaluating} (since different frequencies are reflected in directions different from the central frequency), side-lobe effects~\cite{vordonis2025evaluating}, and other imperfections both in the RIS and the indoor multipath channel~\cite{vordonis2025evaluating}.

\input{plot_LU-ED_SecRate_wro_locs}
Table~\ref{tab:merged_power_full_grouped} also shows the received powers at the LU and ED for the approaches maximizing the LU's and minimizing the ED's rate, respectively. Compared to the SSE-based optimization algorithms, it becomes evident that beamforming towards the LU yields a higher received power at the LU, as it neglects the constraint imposed by the presence of the ED. However, in the context of PLS, it is not sufficient to solely increase the LU’s power; the ED’s power must simultaneously be effectively suppressed. Therefore, SSE-based maximization methods achieve superior secrecy rate performance as compared to those focusing on maximizing the individual achievable rates.

\section{Conclusion}
\label{s:conclusion}
This paper focused on the assessment of the RIS capabilities to enhance PLS through practical indoor measurements. By leveraging a varactor-based dual-polarized RIS prototype tailored for the FR1 band, particularly at~$3.55$~GHz, two distinct algorithms for optimizing the SSE performance were compared: one enhancing the LU’s signal while suppressing its contribution towards the ED position, and another partitioning the RIS to serve intended and unintended users separately. The solutions of these approaches were also compared with conventional ones that optimize the LU's achievable rate. Furthermore, a methodology for designing RIS configuration codebooks in multipath indoor environments was introduced. The presented experimental results not only highlighted the effectiveness of the RIS in securing wireless communications, but also underscored the need to validate RIS designs in practical scenarios, as they hold significant promise for shaping emerging applications for the upcoming 6G networks.

\section{Acknowledgement}
\vspace{-0.12cm}
This work has been supported by the ESA Project PARTICLE: Technological Enablers of Cellular Networks for PVT Assurance (NAVISP-EL1-077). 

\balance
\bibliographystyle{IEEEtran}
\bibliography{IEEEfull,sec_refs}

\end{document}

%% file: confimages/indoor_setting.tex
\begin{tikzpicture}




\draw[thick, dashed] (0,0) ++(-180:1) arc (-180:0:1);
\draw[thick, dashed] (0,0) ++(-180:2) arc (-180:0:2);
\draw[thick, dashed] (0,0) ++(-180:3) arc (-180:0:3);
\draw[thick] (0,0) -- ({3*cos(-105)},{3*sin(-105)});

\node[draw, diamond, minimum size=0.1cm, fill=blue!50, scale=0.43] at ({2*cos(-105)},{2*sin(-105)}) {Tx};

\fill[gray!20] 
    (0,0) -- 
    (-97.5:3) arc[start angle=-97.5, end angle=-82.5, radius=3] -- cycle;
\fill[gray!30] 
    (0,0) -- 
    (-82.5:3) arc[start angle=-82.5, end angle=-67.5, radius=3] -- cycle;
\fill[gray!40] 
    (0,0) -- 
    (-67.5:3) arc[start angle=-67.5, end angle=-52.5, radius=3] -- cycle;
\fill[gray!50] 
    (0,0) -- 
    (-52.5:3) arc[start angle=-52.5, end angle=-37.5, radius=3] -- cycle;
\fill[gray!60] 
    (0,0) -- 
    (-37.5:3) arc[start angle=-37.5, end angle=-22.5, radius=3] -- cycle;
\foreach \angle in {0,-15,...,-90} {
    \draw[thick] (0,0) -- ({3*cos(\angle)},{3*sin(\angle)});
    \pgfmathsetmacro{\tmp}{90 + \angle};
    \pgfmathparse{int(round(\tmp))};
    \let\theIntItmp\pgfmathresult;
    \node[below] at ({3.3*cos(\angle)},{3*sin(\angle)}) {$\theIntItmp \degree$};
}

\foreach \angle in {-60,-75,-90}{
\node[draw, diamond, minimum size=0.1cm, fill=green!50, scale=0.2] at ({3*cos(\angle)},{3*sin(\angle)}) {LU};
}
\foreach \angle in {-45,-60,-75,-90}{
\node[draw, diamond, minimum size=0.1cm, fill=red!50, scale=0.2] at ({2.9*cos(\angle)},{2.9*sin(\angle)}) {ED};
}
\draw[thick] (0,0) -- ({3*cos(-180)},{3*sin(-180)});

\node[draw, rectangle, fill=blue!20, text=black] at (0,0) {RIS};
\end{tikzpicture}

%% file: plot_LU-ED_Rec_power_wro_loc.tex
\begin{table*}[t]
\centering
\caption{LU/ED Received Powers at the LU/ED Locations.}
\label{tab:merged_power_full_grouped} 
\setlength{\tabcolsep}{2pt}
\begin{tabular}{@{}
>{\centering\arraybackslash} m{0.8cm} 
>{\centering\arraybackslash} m{0.8cm} 
S[table-format=-2.2,table-column-width=1cm] 
S[table-format=-2.2,table-column-width=1cm] 
S[table-format=-2.2,table-column-width=1.3cm] 
S[table-format=-2.2,table-column-width=1.3cm] 
S[table-format=-2.2,table-column-width=1.1cm] 
S[table-format=-2.2,table-column-width=1.1cm] 
S[table-format=-2.2,table-column-width=1.35cm] 
S[table-format=-2.2,table-column-width=1.35cm] 
S[table-format=-2.2,table-column-width=1.35cm] 
S[table-format=-2.2,table-column-width=1.35cm] 
S[table-format=-2.2,table-column-width=1cm] 
>{\arraybackslash}S[table-format=-2.2,table-column-width=1cm] 
@{}}
\toprule
\multirow{3}{*}{\parbox[c]{0.8cm}{\centering\textbf{LU\\loc.}}} 
& \multirow{3}{*}{\parbox[c]{0.8cm}{\centering\textbf{ED\\loc.}}} 
  & \multicolumn{6}{c}{\textit{SSE Maximization}} 
  & \multicolumn{4}{c}{\textit{Rate Maximization}} 
  & \multicolumn{2}{c}{\multirow{2}{*}{\parbox[c]{2cm}{\centering \textbf{Uniform\\Surface}}}} \\ 
\cmidrule(lr){3-8} 
\cmidrule(lr){9-12} 
&   & \multicolumn{4}{c}{\textbf{Algorithm~\ref{algo:1}}} 
    & \multicolumn{2}{c}{\textbf{Algorithm~\ref{algo:2}}} 
    & \multicolumn{2}{c}{\textbf{LU max.}}                
    & \multicolumn{2}{c}{\textbf{ED min.}}                
    & \multicolumn{2}{c}{} \\ 
\cmidrule(lr){3-6} 
\cmidrule(lr){7-8} 
\cmidrule(lr){9-10} 
\cmidrule(lr){11-12} 
\cmidrule(l){13-14}
 &  & {LU} & {ED} & {\centering LU-PRS} & {\centering ED-PRS} & {LU} & {ED} & {LU} & {ED} & {LU} & {ED} & {LU} & {ED} \\
 &  & {(dB)} & {(dB)} & {(dB)} & {(dB)} & {(dB)} & {(dB)} & {(dB)} & {(dB)} & {(dB)} & {(dB)} & {(dB)} &{(dB)} \\
\cmidrule{1-2}
\cmidrule(rl){3-4}
\cmidrule(rl){5-6}
\cmidrule(rl){7-8}
\cmidrule(rl){9-10}
\cmidrule(rl){11-12}
\cmidrule(l){13-14}
$0\degree$ & $15\degree$ & -72.85 & -85.47 & -69.66 & -70.16 & -68.25 & -84.70 & -57.57 & -65.10 & -64.36 & -76.82 & -70.86 & -65.96 \\
$0\degree$ & $30\degree$ & -61.76 & -64.40 & -69.84 & -70.19 & -68.15 & -62.42 & -57.85 & -65.03 & -69.92 & -60.74 & -70.55 & -62.95 \\
$0\degree$ & $45\degree$ & -65.11 & -84.74 & -69.71 & -70.14 & -67.66 & -84.18 & -57.24 & -66.70 & -64.00 & -73.96 & -71.03 & -67.94 \\
$15\degree$& $0\degree$  & -65.42 & -84.31 & -69.51 & -70.01 & -60.67 & -71.44 & -56.88 & -59.89 & -69.63 & -75.75 & -66.94 & -70.86 \\
$15\degree$& $30\degree$ & -63.64 & -72.51 & -69.83 & -70.21 & -66.47 & -76.23 & -57.21 & -73.35 & -76.66 & -78.64 & -65.70 & -59.61 \\
$15\degree$& $45\degree$ & -68.67 & -82.44 & -69.21 & -69.71 & -59.42 & -77.65 & -58.02 & -61.40 & -73.64 & -66.15 & -68.58 & -71.76 \\
$30\degree$& $0\degree$  & -62.58 & -80.86 & -69.74 & -70.10 & -63.94 & -75.33 & -56.87 & -63.21 & -77.97 & -74.16 & -63.33 & -72.57 \\
$30\degree$& $15\degree$ & -62.63 & -79.46 & -69.61 & -70.19 & -63.10 & -82.92 & -57.03 & -60.37 & -70.50 & -75.16 & -63.20 & -73.11 \\
$30\degree$& $45\degree$ & -61.81 & -82.30 & -69.74 & -70.14 & -65.99 & -82.73 & -56.10 & -74.77 & -64.64 & -76.87 & -63.13 & -67.84 \\
\bottomrule
\end{tabular}
\end{table*}

%% file: heatmap.tex
\begin{figure}[tb] 
\centering 
    \centering
    \scalebox{0.51}{\input{sec_heatmap.tex}} 
\caption{Achievable SSE (bits/s/Hz) with all implemented RIS design approaches, each in a different region as shown in the top-left gray-colored inset. The LU/ED locations were defined as depicted in Fig.~\ref{fig:Channel_Model2}. Note that the LU and ED were not positioned at the same azimuth angle simultaneously.}
\label{fig:combined_sec_cap} 
\end{figure}
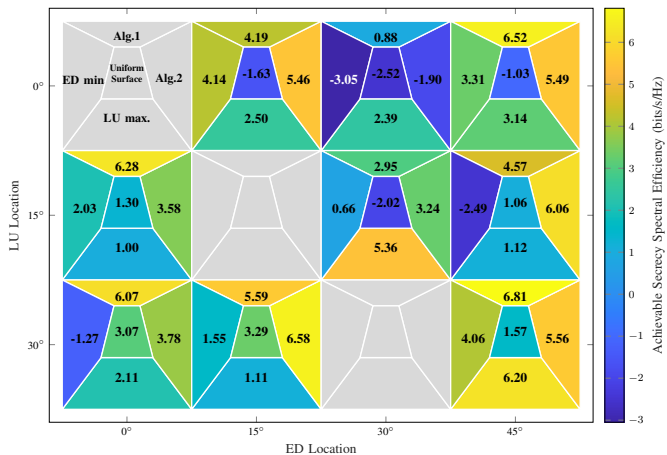

%% file: sec_heatmap.tex
%
%
\definecolor{mycolor1}{rgb}{0.20610,0.78080,0.60650}%
\definecolor{mycolor2}{rgb}{0.99660,0.77400,0.21380}%
\definecolor{mycolor3}{rgb}{0.73440,0.76790,0.18520}%
\definecolor{mycolor4}{rgb}{0.72180,0.77030,0.19240}%
\definecolor{mycolor5}{rgb}{0.28090,0.32590,0.96000}%
\definecolor{mycolor6}{rgb}{0.19050,0.77430,0.63030}%
\definecolor{mycolor7}{rgb}{0.28090,0.28750,0.93050}%
\definecolor{mycolor8}{rgb}{0.10850,0.66690,0.87340}%
\definecolor{mycolor9}{rgb}{0.24220,0.15040,0.66030}%
\definecolor{mycolor10}{rgb}{0.26760,0.20520,0.81480}%
\definecolor{mycolor11}{rgb}{0.36710,0.80210,0.45630}%
\definecolor{mycolor12}{rgb}{0.99620,0.77980,0.20950}%
\definecolor{mycolor13}{rgb}{0.96420,0.94370,0.12160}%
\definecolor{mycolor14}{rgb}{0.41840,0.80300,0.41220}%
\definecolor{mycolor15}{rgb}{0.26210,0.40880,0.99460}%
\definecolor{mycolor16}{rgb}{0.09640,0.67890,0.86090}%
\definecolor{mycolor17}{rgb}{0.51870,0.79820,0.33740}%
\definecolor{mycolor18}{rgb}{0.95960,0.90230,0.14800}%
\definecolor{mycolor19}{rgb}{0.11410,0.75310,0.70190}%
\definecolor{mycolor20}{rgb}{0.03280,0.70710,0.81830}%
\definecolor{mycolor21}{rgb}{0.99710,0.76260,0.22240}%
\definecolor{mycolor22}{rgb}{0.40500,0.80310,0.42330}%
\definecolor{mycolor23}{rgb}{0.30520,0.79850,0.50740}%
\definecolor{mycolor24}{rgb}{0.12880,0.64080,0.89100}%
\definecolor{mycolor25}{rgb}{0.27980,0.27080,0.91500}%
\definecolor{mycolor26}{rgb}{0.07890,0.69020,0.84620}%
\definecolor{mycolor27}{rgb}{0.96510,0.87160,0.16080}%
\definecolor{mycolor28}{rgb}{0.84950,0.74350,0.15640}%
\definecolor{mycolor29}{rgb}{0.09140,0.68280,0.85620}%
\definecolor{mycolor30}{rgb}{0.13540,0.75720,0.68810}%
\definecolor{mycolor31}{rgb}{0.60240,0.78960,0.27260}%
\definecolor{mycolor32}{rgb}{0.27410,0.37480,0.98620}%
\definecolor{mycolor33}{rgb}{0.34240,0.80090,0.47740}%
\definecolor{mycolor34}{rgb}{0.96570,0.94940,0.11680}%
\definecolor{mycolor35}{rgb}{0.99490,0.79150,0.20120}%
\definecolor{mycolor36}{rgb}{0.00090,0.72480,0.78150}%
\definecolor{mycolor37}{rgb}{0.96080,0.89020,0.15320}%
\definecolor{mycolor38}{rgb}{0.97690,0.98390,0.08050}%
\definecolor{mycolor39}{rgb}{0.69630,0.77500,0.20750}%
\begin{tikzpicture}

\begin{axis}[%
width=5.581in,
height=4.252in,
at={(0.936in,0.839in)},
scale only axis,
point meta min=-3.05,
point meta max=6.81,
xmin=-0.1,
xmax=4.1,
xtick={0.5,1.5,2.5,3.5},
xticklabels={{$\text{0}^\circ$},{$\text{15}^\circ$},{$\text{30}^\circ$},{$\text{45}^\circ$}},
xlabel style={font=\color{white!15!black}},
xlabel={ED Location},
y dir=reverse,
ymin=-0.1,
ymax=3.1,
ytick={0.5,1.5,2.5},
yticklabels={{$\text{0}^\circ$},{$\text{15}^\circ$},{$\text{30}^\circ$}},
ylabel style={font=\color{white!15!black}},
ylabel={LU Location},
axis background/.style={fill=white},
colormap={mymap}{[1pt] rgb(0pt)=(0.2422,0.1504,0.6603); rgb(1pt)=(0.2444,0.1534,0.6728); rgb(2pt)=(0.2464,0.1569,0.6847); rgb(3pt)=(0.2484,0.1607,0.6961); rgb(4pt)=(0.2503,0.1648,0.7071); rgb(5pt)=(0.2522,0.1689,0.7179); rgb(6pt)=(0.254,0.1732,0.7286); rgb(7pt)=(0.2558,0.1773,0.7393); rgb(8pt)=(0.2576,0.1814,0.7501); rgb(9pt)=(0.2594,0.1854,0.761); rgb(11pt)=(0.2628,0.1932,0.7828); rgb(12pt)=(0.2645,0.1972,0.7937); rgb(13pt)=(0.2661,0.2011,0.8043); rgb(14pt)=(0.2676,0.2052,0.8148); rgb(15pt)=(0.2691,0.2094,0.8249); rgb(16pt)=(0.2704,0.2138,0.8346); rgb(17pt)=(0.2717,0.2184,0.8439); rgb(18pt)=(0.2729,0.2231,0.8528); rgb(19pt)=(0.274,0.228,0.8612); rgb(20pt)=(0.2749,0.233,0.8692); rgb(21pt)=(0.2758,0.2382,0.8767); rgb(22pt)=(0.2766,0.2435,0.884); rgb(23pt)=(0.2774,0.2489,0.8908); rgb(24pt)=(0.2781,0.2543,0.8973); rgb(25pt)=(0.2788,0.2598,0.9035); rgb(26pt)=(0.2794,0.2653,0.9094); rgb(27pt)=(0.2798,0.2708,0.915); rgb(28pt)=(0.2802,0.2764,0.9204); rgb(29pt)=(0.2806,0.2819,0.9255); rgb(30pt)=(0.2809,0.2875,0.9305); rgb(31pt)=(0.2811,0.293,0.9352); rgb(32pt)=(0.2813,0.2985,0.9397); rgb(33pt)=(0.2814,0.304,0.9441); rgb(34pt)=(0.2814,0.3095,0.9483); rgb(35pt)=(0.2813,0.315,0.9524); rgb(36pt)=(0.2811,0.3204,0.9563); rgb(37pt)=(0.2809,0.3259,0.96); rgb(38pt)=(0.2807,0.3313,0.9636); rgb(39pt)=(0.2803,0.3367,0.967); rgb(40pt)=(0.2798,0.3421,0.9702); rgb(41pt)=(0.2791,0.3475,0.9733); rgb(42pt)=(0.2784,0.3529,0.9763); rgb(43pt)=(0.2776,0.3583,0.9791); rgb(44pt)=(0.2766,0.3638,0.9817); rgb(45pt)=(0.2754,0.3693,0.984); rgb(46pt)=(0.2741,0.3748,0.9862); rgb(47pt)=(0.2726,0.3804,0.9881); rgb(48pt)=(0.271,0.386,0.9898); rgb(49pt)=(0.2691,0.3916,0.9912); rgb(50pt)=(0.267,0.3973,0.9924); rgb(51pt)=(0.2647,0.403,0.9935); rgb(52pt)=(0.2621,0.4088,0.9946); rgb(53pt)=(0.2591,0.4145,0.9955); rgb(54pt)=(0.2556,0.4203,0.9965); rgb(55pt)=(0.2517,0.4261,0.9974); rgb(56pt)=(0.2473,0.4319,0.9983); rgb(57pt)=(0.2424,0.4378,0.9991); rgb(58pt)=(0.2369,0.4437,0.9996); rgb(59pt)=(0.2311,0.4497,0.9995); rgb(60pt)=(0.225,0.4559,0.9985); rgb(61pt)=(0.2189,0.462,0.9968); rgb(62pt)=(0.2128,0.4682,0.9948); rgb(63pt)=(0.2066,0.4743,0.9926); rgb(64pt)=(0.2006,0.4803,0.9906); rgb(65pt)=(0.195,0.4861,0.9887); rgb(66pt)=(0.1903,0.4919,0.9867); rgb(67pt)=(0.1869,0.4975,0.9844); rgb(68pt)=(0.1847,0.503,0.9819); rgb(69pt)=(0.1831,0.5084,0.9793); rgb(70pt)=(0.1818,0.5138,0.9766); rgb(71pt)=(0.1806,0.5191,0.9738); rgb(72pt)=(0.1795,0.5244,0.9709); rgb(73pt)=(0.1785,0.5296,0.9677); rgb(74pt)=(0.1778,0.5349,0.9641); rgb(75pt)=(0.1773,0.5401,0.9602); rgb(76pt)=(0.1768,0.5452,0.956); rgb(77pt)=(0.1764,0.5504,0.9516); rgb(78pt)=(0.1755,0.5554,0.9473); rgb(79pt)=(0.174,0.5605,0.9432); rgb(80pt)=(0.1716,0.5655,0.9393); rgb(81pt)=(0.1686,0.5705,0.9357); rgb(82pt)=(0.1649,0.5755,0.9323); rgb(83pt)=(0.161,0.5805,0.9289); rgb(84pt)=(0.1573,0.5854,0.9254); rgb(85pt)=(0.154,0.5902,0.9218); rgb(86pt)=(0.1513,0.595,0.9182); rgb(87pt)=(0.1492,0.5997,0.9147); rgb(88pt)=(0.1475,0.6043,0.9113); rgb(89pt)=(0.1461,0.6089,0.908); rgb(90pt)=(0.1446,0.6135,0.905); rgb(91pt)=(0.1429,0.618,0.9022); rgb(92pt)=(0.1408,0.6226,0.8998); rgb(93pt)=(0.1383,0.6272,0.8975); rgb(94pt)=(0.1354,0.6317,0.8953); rgb(95pt)=(0.1321,0.6363,0.8932); rgb(96pt)=(0.1288,0.6408,0.891); rgb(97pt)=(0.1253,0.6453,0.8887); rgb(98pt)=(0.1219,0.6497,0.8862); rgb(99pt)=(0.1185,0.6541,0.8834); rgb(100pt)=(0.1152,0.6584,0.8804); rgb(101pt)=(0.1119,0.6627,0.877); rgb(102pt)=(0.1085,0.6669,0.8734); rgb(103pt)=(0.1048,0.671,0.8695); rgb(104pt)=(0.1009,0.675,0.8653); rgb(105pt)=(0.0964,0.6789,0.8609); rgb(106pt)=(0.0914,0.6828,0.8562); rgb(107pt)=(0.0855,0.6865,0.8513); rgb(108pt)=(0.0789,0.6902,0.8462); rgb(109pt)=(0.0713,0.6938,0.8409); rgb(110pt)=(0.0628,0.6972,0.8355); rgb(111pt)=(0.0535,0.7006,0.8299); rgb(112pt)=(0.0433,0.7039,0.8242); rgb(113pt)=(0.0328,0.7071,0.8183); rgb(114pt)=(0.0234,0.7103,0.8124); rgb(115pt)=(0.0155,0.7133,0.8064); rgb(116pt)=(0.0091,0.7163,0.8003); rgb(117pt)=(0.0046,0.7192,0.7941); rgb(118pt)=(0.0019,0.722,0.7878); rgb(119pt)=(0.0009,0.7248,0.7815); rgb(120pt)=(0.0018,0.7275,0.7752); rgb(121pt)=(0.0046,0.7301,0.7688); rgb(122pt)=(0.0094,0.7327,0.7623); rgb(123pt)=(0.0162,0.7352,0.7558); rgb(124pt)=(0.0253,0.7376,0.7492); rgb(125pt)=(0.0369,0.74,0.7426); rgb(126pt)=(0.0504,0.7423,0.7359); rgb(127pt)=(0.0638,0.7446,0.7292); rgb(128pt)=(0.077,0.7468,0.7224); rgb(129pt)=(0.0899,0.7489,0.7156); rgb(130pt)=(0.1023,0.751,0.7088); rgb(131pt)=(0.1141,0.7531,0.7019); rgb(132pt)=(0.1252,0.7552,0.695); rgb(133pt)=(0.1354,0.7572,0.6881); rgb(134pt)=(0.1448,0.7593,0.6812); rgb(135pt)=(0.1532,0.7614,0.6741); rgb(136pt)=(0.1609,0.7635,0.6671); rgb(137pt)=(0.1678,0.7656,0.6599); rgb(138pt)=(0.1741,0.7678,0.6527); rgb(139pt)=(0.1799,0.7699,0.6454); rgb(140pt)=(0.1853,0.7721,0.6379); rgb(141pt)=(0.1905,0.7743,0.6303); rgb(142pt)=(0.1954,0.7765,0.6225); rgb(143pt)=(0.2003,0.7787,0.6146); rgb(144pt)=(0.2061,0.7808,0.6065); rgb(145pt)=(0.2118,0.7828,0.5983); rgb(146pt)=(0.2178,0.7849,0.5899); rgb(147pt)=(0.2244,0.7869,0.5813); rgb(148pt)=(0.2318,0.7887,0.5725); rgb(149pt)=(0.2401,0.7905,0.5636); rgb(150pt)=(0.2491,0.7922,0.5546); rgb(151pt)=(0.2589,0.7937,0.5454); rgb(152pt)=(0.2695,0.7951,0.536); rgb(153pt)=(0.2809,0.7964,0.5266); rgb(154pt)=(0.2929,0.7975,0.517); rgb(155pt)=(0.3052,0.7985,0.5074); rgb(156pt)=(0.3176,0.7994,0.4975); rgb(157pt)=(0.3301,0.8002,0.4876); rgb(158pt)=(0.3424,0.8009,0.4774); rgb(159pt)=(0.3548,0.8016,0.4669); rgb(160pt)=(0.3671,0.8021,0.4563); rgb(161pt)=(0.3795,0.8026,0.4454); rgb(162pt)=(0.3921,0.8029,0.4344); rgb(163pt)=(0.405,0.8031,0.4233); rgb(164pt)=(0.4184,0.803,0.4122); rgb(165pt)=(0.4322,0.8028,0.4013); rgb(166pt)=(0.4463,0.8024,0.3904); rgb(167pt)=(0.4608,0.8018,0.3797); rgb(168pt)=(0.4753,0.8011,0.3691); rgb(169pt)=(0.4899,0.8002,0.3586); rgb(170pt)=(0.5044,0.7993,0.348); rgb(171pt)=(0.5187,0.7982,0.3374); rgb(172pt)=(0.5329,0.797,0.3267); rgb(173pt)=(0.547,0.7957,0.3159); rgb(175pt)=(0.5748,0.7929,0.2941); rgb(176pt)=(0.5886,0.7913,0.2833); rgb(177pt)=(0.6024,0.7896,0.2726); rgb(178pt)=(0.6161,0.7878,0.2622); rgb(179pt)=(0.6297,0.7859,0.2521); rgb(180pt)=(0.6433,0.7839,0.2423); rgb(181pt)=(0.6567,0.7818,0.2329); rgb(182pt)=(0.6701,0.7796,0.2239); rgb(183pt)=(0.6833,0.7773,0.2155); rgb(184pt)=(0.6963,0.775,0.2075); rgb(185pt)=(0.7091,0.7727,0.1998); rgb(186pt)=(0.7218,0.7703,0.1924); rgb(187pt)=(0.7344,0.7679,0.1852); rgb(188pt)=(0.7468,0.7654,0.1782); rgb(189pt)=(0.759,0.7629,0.1717); rgb(190pt)=(0.771,0.7604,0.1658); rgb(191pt)=(0.7829,0.7579,0.1608); rgb(192pt)=(0.7945,0.7554,0.157); rgb(193pt)=(0.806,0.7529,0.1546); rgb(194pt)=(0.8172,0.7505,0.1535); rgb(195pt)=(0.8281,0.7481,0.1536); rgb(196pt)=(0.8389,0.7457,0.1546); rgb(197pt)=(0.8495,0.7435,0.1564); rgb(198pt)=(0.86,0.7413,0.1587); rgb(199pt)=(0.8703,0.7392,0.1615); rgb(200pt)=(0.8804,0.7372,0.165); rgb(201pt)=(0.8903,0.7353,0.1695); rgb(202pt)=(0.9,0.7336,0.1749); rgb(203pt)=(0.9093,0.7321,0.1815); rgb(204pt)=(0.9184,0.7308,0.189); rgb(205pt)=(0.9272,0.7298,0.1973); rgb(206pt)=(0.9357,0.729,0.2061); rgb(207pt)=(0.944,0.7285,0.2151); rgb(208pt)=(0.9523,0.7284,0.2237); rgb(209pt)=(0.9606,0.7285,0.2312); rgb(210pt)=(0.9689,0.7292,0.2373); rgb(211pt)=(0.977,0.7304,0.2418); rgb(212pt)=(0.9842,0.733,0.2446); rgb(213pt)=(0.99,0.7365,0.2429); rgb(214pt)=(0.9946,0.7407,0.2394); rgb(215pt)=(0.9966,0.7458,0.2351); rgb(216pt)=(0.9971,0.7513,0.2309); rgb(217pt)=(0.9972,0.7569,0.2267); rgb(218pt)=(0.9971,0.7626,0.2224); rgb(219pt)=(0.9969,0.7683,0.2181); rgb(220pt)=(0.9966,0.774,0.2138); rgb(221pt)=(0.9962,0.7798,0.2095); rgb(222pt)=(0.9957,0.7856,0.2053); rgb(223pt)=(0.9949,0.7915,0.2012); rgb(224pt)=(0.9938,0.7974,0.1974); rgb(225pt)=(0.9923,0.8034,0.1939); rgb(226pt)=(0.9906,0.8095,0.1906); rgb(227pt)=(0.9885,0.8156,0.1875); rgb(228pt)=(0.9861,0.8218,0.1846); rgb(229pt)=(0.9835,0.828,0.1817); rgb(230pt)=(0.9807,0.8342,0.1787); rgb(231pt)=(0.9778,0.8404,0.1757); rgb(232pt)=(0.9748,0.8467,0.1726); rgb(233pt)=(0.972,0.8529,0.1695); rgb(234pt)=(0.9694,0.8591,0.1665); rgb(235pt)=(0.9671,0.8654,0.1636); rgb(236pt)=(0.9651,0.8716,0.1608); rgb(237pt)=(0.9634,0.8778,0.1582); rgb(238pt)=(0.9619,0.884,0.1557); rgb(239pt)=(0.9608,0.8902,0.1532); rgb(240pt)=(0.9601,0.8963,0.1507); rgb(241pt)=(0.9596,0.9023,0.148); rgb(242pt)=(0.9595,0.9084,0.145); rgb(243pt)=(0.9597,0.9143,0.1418); rgb(244pt)=(0.9601,0.9203,0.1382); rgb(245pt)=(0.9608,0.9262,0.1344); rgb(246pt)=(0.9618,0.932,0.1304); rgb(247pt)=(0.9629,0.9379,0.1261); rgb(248pt)=(0.9642,0.9437,0.1216); rgb(249pt)=(0.9657,0.9494,0.1168); rgb(250pt)=(0.9674,0.9552,0.1116); rgb(251pt)=(0.9692,0.9609,0.1061); rgb(252pt)=(0.9711,0.9667,0.1001); rgb(253pt)=(0.973,0.9724,0.0938); rgb(254pt)=(0.9749,0.9782,0.0872); rgb(255pt)=(0.9769,0.9839,0.0805)},
colorbar,
colorbar style={ylabel style={font=\color{white!15!black}}, ylabel={Achievable Secrecy Spectral Efficiency (bits/s/Hz)}}
]

\addplot[area legend, line width=1.0pt, draw=white, fill=white!85!black, forget plot]
table[row sep=crcr] {%
x	y\\
0	1\\
1	1\\
0.7	0.6\\
0.3	0.6\\
}--cycle;
\node[centered, align=center, inner sep=0, font=\bfseries\small]
at (axis cs:0.5,0.75) {LU max.};

\addplot[area legend, line width=1.0pt, draw=white, fill=white!85!black, forget plot]
table[row sep=crcr] {%
x	y\\
1	1\\
1	0\\
0.6	0.2\\
0.7	0.6\\
}--cycle;
\node[centered, align=center, inner sep=0, font=\bfseries\small]
at (axis cs:0.825,0.45) {Alg.2};

\addplot[area legend, line width=1.0pt, draw=white, fill=white!85!black, forget plot]
table[row sep=crcr] {%
x	y\\
0.6	0.2\\
1	0\\
0	0\\
0.4	0.2\\
}--cycle;
\node[centered, align=center, inner sep=0, font=\bfseries\small]
at (axis cs:0.5,0.125) {Alg.1};

\addplot[area legend, line width=1.0pt, draw=white, fill=white!85!black, forget plot]
table[row sep=crcr] {%
x	y\\
0	1\\
0	0\\
0.4	0.2\\
0.3	0.6\\
}--cycle;
\node[centered, align=center, inner sep=0, font=\bfseries\small]
at (axis cs:0.175,0.45) {ED min.};

\addplot[area legend, line width=1.0pt, draw=white, fill=white!85!black, forget plot]
table[row sep=crcr] {%
x	y\\
0.3	0.6\\
0.7	0.6\\
0.6	0.2\\
0.4	0.2\\
}--cycle;
\node[centered, align=center, inner sep=0, font=\bfseries\scriptsize]
at (axis cs:0.5,0.4) {Uniform\\Surface};

\addplot[area legend, line width=1.0pt, draw=white, fill=mycolor1, forget plot]
table[row sep=crcr] {%
x	y\\
1	1\\
2	1\\
1.7	0.6\\
1.3	0.6\\
}--cycle;
\node[centered, align=center, inner sep=0, font=\bfseries]
at (axis cs:1.5,0.75) {2.50};

\addplot[area legend, line width=1.0pt, draw=white, fill=mycolor2, forget plot]
table[row sep=crcr] {%
x	y\\
2	1\\
2	0\\
1.6	0.2\\
1.7	0.6\\
}--cycle;
\node[centered, align=center, inner sep=0, font=\bfseries]
at (axis cs:1.825,0.45) {5.46};

\addplot[area legend, line width=1.0pt, draw=white, fill=mycolor3, forget plot]
table[row sep=crcr] {%
x	y\\
1.6	0.2\\
2	0\\
1	0\\
1.4	0.2\\
}--cycle;
\node[centered, align=center, inner sep=0, font=\bfseries]
at (axis cs:1.5,0.125) {4.19};

\addplot[area legend, line width=1.0pt, draw=white, fill=mycolor4, forget plot]
table[row sep=crcr] {%
x	y\\
1	1\\
1	0\\
1.4	0.2\\
1.3	0.6\\
}--cycle;
\node[centered, align=center, inner sep=0, font=\bfseries]
at (axis cs:1.175,0.45) {4.14};

\addplot[area legend, line width=1.0pt, draw=white, fill=mycolor5, forget plot]
table[row sep=crcr] {%
x	y\\
1.3	0.6\\
1.7	0.6\\
1.6	0.2\\
1.4	0.2\\
}--cycle;
\node[centered, align=center, inner sep=0, font=\bfseries]
at (axis cs:1.5,0.4) {-1.63};

\addplot[area legend, line width=1.0pt, draw=white, fill=mycolor6, forget plot]
table[row sep=crcr] {%
x	y\\
2	1\\
3	1\\
2.7	0.6\\
2.3	0.6\\
}--cycle;
\node[centered, align=center, inner sep=0, font=\bfseries]
at (axis cs:2.5,0.75) {2.39};

\addplot[area legend, line width=1.0pt, draw=white, fill=mycolor7, forget plot]
table[row sep=crcr] {%
x	y\\
3	1\\
3	0\\
2.6	0.2\\
2.7	0.6\\
}--cycle;
\node[centered, align=center, inner sep=0, font=\bfseries]
at (axis cs:2.825,0.45) {-1.90};

\addplot[area legend, line width=1.0pt, draw=white, fill=mycolor8, forget plot]
table[row sep=crcr] {%
x	y\\
2.6	0.2\\
3	0\\
2	0\\
2.4	0.2\\
}--cycle;
\node[centered, align=center, inner sep=0, font=\bfseries]
at (axis cs:2.5,0.125) {0.88};

\addplot[area legend, line width=1.0pt, draw=white, fill=mycolor9, forget plot]
table[row sep=crcr] {%
x	y\\
2	1\\
2	0\\
2.4	0.2\\
2.3	0.6\\
}--cycle;
\node[centered, align=center, inner sep=0, font=\bfseries\color{white}]
at (axis cs:2.175,0.45) {-3.05};

\addplot[area legend, line width=1.0pt, draw=white, fill=mycolor10, forget plot]
table[row sep=crcr] {%
x	y\\
2.3	0.6\\
2.7	0.6\\
2.6	0.2\\
2.4	0.2\\
}--cycle;
\node[centered, align=center, inner sep=0, font=\bfseries]
at (axis cs:2.5,0.4) {-2.52};

\addplot[area legend, line width=1.0pt, draw=white, fill=mycolor11, forget plot]
table[row sep=crcr] {%
x	y\\
3	1\\
4	1\\
3.7	0.6\\
3.3	0.6\\
}--cycle;
\node[centered, align=center, inner sep=0, font=\bfseries]
at (axis cs:3.5,0.75) {3.14};

\addplot[area legend, line width=1.0pt, draw=white, fill=mycolor12, forget plot]
table[row sep=crcr] {%
x	y\\
4	1\\
4	0\\
3.6	0.2\\
3.7	0.6\\
}--cycle;
\node[centered, align=center, inner sep=0, font=\bfseries]
at (axis cs:3.825,0.45) {5.49};

\addplot[area legend, line width=1.0pt, draw=white, fill=mycolor13, forget plot]
table[row sep=crcr] {%
x	y\\
3.6	0.2\\
4	0\\
3	0\\
3.4	0.2\\
}--cycle;
\node[centered, align=center, inner sep=0, font=\bfseries]
at (axis cs:3.5,0.125) {6.52};

\addplot[area legend, line width=1.0pt, draw=white, fill=mycolor14, forget plot]
table[row sep=crcr] {%
x	y\\
3	1\\
3	0\\
3.4	0.2\\
3.3	0.6\\
}--cycle;
\node[centered, align=center, inner sep=0, font=\bfseries]
at (axis cs:3.175,0.45) {3.31};

\addplot[area legend, line width=1.0pt, draw=white, fill=mycolor15, forget plot]
table[row sep=crcr] {%
x	y\\
3.3	0.6\\
3.7	0.6\\
3.6	0.2\\
3.4	0.2\\
}--cycle;
\node[centered, align=center, inner sep=0, font=\bfseries]
at (axis cs:3.5,0.4) {-1.03};

\addplot[area legend, line width=1.0pt, draw=white, fill=mycolor16, forget plot]
table[row sep=crcr] {%
x	y\\
0	2\\
1	2\\
0.7	1.6\\
0.3	1.6\\
}--cycle;
\node[centered, align=center, inner sep=0, font=\bfseries]
at (axis cs:0.5,1.75) {1.00};

\addplot[area legend, line width=1.0pt, draw=white, fill=mycolor17, forget plot]
table[row sep=crcr] {%
x	y\\
1	2\\
1	1\\
0.6	1.2\\
0.7	1.6\\
}--cycle;
\node[centered, align=center, inner sep=0, font=\bfseries]
at (axis cs:0.825,1.45) {3.58};

\addplot[area legend, line width=1.0pt, draw=white, fill=mycolor18, forget plot]
table[row sep=crcr] {%
x	y\\
0.6	1.2\\
1	1\\
0	1\\
0.4	1.2\\
}--cycle;
\node[centered, align=center, inner sep=0, font=\bfseries]
at (axis cs:0.5,1.125) {6.28};

\addplot[area legend, line width=1.0pt, draw=white, fill=mycolor19, forget plot]
table[row sep=crcr] {%
x	y\\
0	2\\
0	1\\
0.4	1.2\\
0.3	1.6\\
}--cycle;
\node[centered, align=center, inner sep=0, font=\bfseries]
at (axis cs:0.175,1.45) {2.03};

\addplot[area legend, line width=1.0pt, draw=white, fill=mycolor20, forget plot]
table[row sep=crcr] {%
x	y\\
0.3	1.6\\
0.7	1.6\\
0.6	1.2\\
0.4	1.2\\
}--cycle;
\node[centered, align=center, inner sep=0, font=\bfseries]
at (axis cs:0.5,1.4) {1.30};

\addplot[area legend, line width=1.0pt, draw=white, fill=white!85!black, forget plot]
table[row sep=crcr] {%
x	y\\
1	2\\
2	2\\
1.7	1.6\\
1.3	1.6\\
}--cycle;

\addplot[area legend, line width=1.0pt, draw=white, fill=white!85!black, forget plot]
table[row sep=crcr] {%
x	y\\
2	2\\
2	1\\
1.6	1.2\\
1.7	1.6\\
}--cycle;

\addplot[area legend, line width=1.0pt, draw=white, fill=white!85!black, forget plot]
table[row sep=crcr] {%
x	y\\
1.6	1.2\\
2	1\\
1	1\\
1.4	1.2\\
}--cycle;

\addplot[area legend, line width=1.0pt, draw=white, fill=white!85!black, forget plot]
table[row sep=crcr] {%
x	y\\
1	2\\
1	1\\
1.4	1.2\\
1.3	1.6\\
}--cycle;

\addplot[area legend, line width=1.0pt, draw=white, fill=white!85!black, forget plot]
table[row sep=crcr] {%
x	y\\
1.3	1.6\\
1.7	1.6\\
1.6	1.2\\
1.4	1.2\\
}--cycle;

\addplot[area legend, line width=1.0pt, draw=white, fill=mycolor21, forget plot]
table[row sep=crcr] {%
x	y\\
2	2\\
3	2\\
2.7	1.6\\
2.3	1.6\\
}--cycle;
\node[centered, align=center, inner sep=0, font=\bfseries]
at (axis cs:2.5,1.75) {5.36};

\addplot[area legend, line width=1.0pt, draw=white, fill=mycolor22, forget plot]
table[row sep=crcr] {%
x	y\\
3	2\\
3	1\\
2.6	1.2\\
2.7	1.6\\
}--cycle;
\node[centered, align=center, inner sep=0, font=\bfseries]
at (axis cs:2.825,1.45) {3.24};

\addplot[area legend, line width=1.0pt, draw=white, fill=mycolor23, forget plot]
table[row sep=crcr] {%
x	y\\
2.6	1.2\\
3	1\\
2	1\\
2.4	1.2\\
}--cycle;
\node[centered, align=center, inner sep=0, font=\bfseries]
at (axis cs:2.5,1.125) {2.95};

\addplot[area legend, line width=1.0pt, draw=white, fill=mycolor24, forget plot]
table[row sep=crcr] {%
x	y\\
2	2\\
2	1\\
2.4	1.2\\
2.3	1.6\\
}--cycle;
\node[centered, align=center, inner sep=0, font=\bfseries]
at (axis cs:2.175,1.45) {0.66};

\addplot[area legend, line width=1.0pt, draw=white, fill=mycolor25, forget plot]
table[row sep=crcr] {%
x	y\\
2.3	1.6\\
2.7	1.6\\
2.6	1.2\\
2.4	1.2\\
}--cycle;
\node[centered, align=center, inner sep=0, font=\bfseries]
at (axis cs:2.5,1.4) {-2.02};

\addplot[area legend, line width=1.0pt, draw=white, fill=mycolor26, forget plot]
table[row sep=crcr] {%
x	y\\
3	2\\
4	2\\
3.7	1.6\\
3.3	1.6\\
}--cycle;
\node[centered, align=center, inner sep=0, font=\bfseries]
at (axis cs:3.5,1.75) {1.12};

\addplot[area legend, line width=1.0pt, draw=white, fill=mycolor27, forget plot]
table[row sep=crcr] {%
x	y\\
4	2\\
4	1\\
3.6	1.2\\
3.7	1.6\\
}--cycle;
\node[centered, align=center, inner sep=0, font=\bfseries]
at (axis cs:3.825,1.45) {6.06};

\addplot[area legend, line width=1.0pt, draw=white, fill=mycolor28, forget plot]
table[row sep=crcr] {%
x	y\\
3.6	1.2\\
4	1\\
3	1\\
3.4	1.2\\
}--cycle;
\node[centered, align=center, inner sep=0, font=\bfseries]
at (axis cs:3.5,1.125) {4.57};

\addplot[area legend, line width=1.0pt, draw=white, fill=mycolor10, forget plot]
table[row sep=crcr] {%
x	y\\
3	2\\
3	1\\
3.4	1.2\\
3.3	1.6\\
}--cycle;
\node[centered, align=center, inner sep=0, font=\bfseries]
at (axis cs:3.175,1.45) {-2.49};

\addplot[area legend, line width=1.0pt, draw=white, fill=mycolor29, forget plot]
table[row sep=crcr] {%
x	y\\
3.3	1.6\\
3.7	1.6\\
3.6	1.2\\
3.4	1.2\\
}--cycle;
\node[centered, align=center, inner sep=0, font=\bfseries]
at (axis cs:3.5,1.4) {1.06};

\addplot[area legend, line width=1.0pt, draw=white, fill=mycolor30, forget plot]
table[row sep=crcr] {%
x	y\\
0	3\\
1	3\\
0.7	2.6\\
0.3	2.6\\
}--cycle;
\node[centered, align=center, inner sep=0, font=\bfseries]
at (axis cs:0.5,2.75) {2.11};

\addplot[area legend, line width=1.0pt, draw=white, fill=mycolor31, forget plot]
table[row sep=crcr] {%
x	y\\
1	3\\
1	2\\
0.6	2.2\\
0.7	2.6\\
}--cycle;
\node[centered, align=center, inner sep=0, font=\bfseries]
at (axis cs:0.825,2.45) {3.78};

\addplot[area legend, line width=1.0pt, draw=white, fill=mycolor27, forget plot]
table[row sep=crcr] {%
x	y\\
0.6	2.2\\
1	2\\
0	2\\
0.4	2.2\\
}--cycle;
\node[centered, align=center, inner sep=0, font=\bfseries]
at (axis cs:0.5,2.125) {6.07};

\addplot[area legend, line width=1.0pt, draw=white, fill=mycolor32, forget plot]
table[row sep=crcr] {%
x	y\\
0	3\\
0	2\\
0.4	2.2\\
0.3	2.6\\
}--cycle;
\node[centered, align=center, inner sep=0, font=\bfseries]
at (axis cs:0.175,2.45) {-1.27};

\addplot[area legend, line width=1.0pt, draw=white, fill=mycolor33, forget plot]
table[row sep=crcr] {%
x	y\\
0.3	2.6\\
0.7	2.6\\
0.6	2.2\\
0.4	2.2\\
}--cycle;
\node[centered, align=center, inner sep=0, font=\bfseries]
at (axis cs:0.5,2.4) {3.07};

\addplot[area legend, line width=1.0pt, draw=white, fill=mycolor26, forget plot]
table[row sep=crcr] {%
x	y\\
1	3\\
2	3\\
1.7	2.6\\
1.3	2.6\\
}--cycle;
\node[centered, align=center, inner sep=0, font=\bfseries]
at (axis cs:1.5,2.75) {1.11};

\addplot[area legend, line width=1.0pt, draw=white, fill=mycolor34, forget plot]
table[row sep=crcr] {%
x	y\\
2	3\\
2	2\\
1.6	2.2\\
1.7	2.6\\
}--cycle;
\node[centered, align=center, inner sep=0, font=\bfseries]
at (axis cs:1.825,2.45) {6.58};

\addplot[area legend, line width=1.0pt, draw=white, fill=mycolor35, forget plot]
table[row sep=crcr] {%
x	y\\
1.6	2.2\\
2	2\\
1	2\\
1.4	2.2\\
}--cycle;
\node[centered, align=center, inner sep=0, font=\bfseries]
at (axis cs:1.5,2.125) {5.59};

\addplot[area legend, line width=1.0pt, draw=white, fill=mycolor36, forget plot]
table[row sep=crcr] {%
x	y\\
1	3\\
1	2\\
1.4	2.2\\
1.3	2.6\\
}--cycle;
\node[centered, align=center, inner sep=0, font=\bfseries]
at (axis cs:1.175,2.45) {1.55};

\addplot[area legend, line width=1.0pt, draw=white, fill=mycolor14, forget plot]
table[row sep=crcr] {%
x	y\\
1.3	2.6\\
1.7	2.6\\
1.6	2.2\\
1.4	2.2\\
}--cycle;
\node[centered, align=center, inner sep=0, font=\bfseries]
at (axis cs:1.5,2.4) {3.29};

\addplot[area legend, line width=1.0pt, draw=white, fill=white!85!black, forget plot]
table[row sep=crcr] {%
x	y\\
2	3\\
3	3\\
2.7	2.6\\
2.3	2.6\\
}--cycle;

\addplot[area legend, line width=1.0pt, draw=white, fill=white!85!black, forget plot]
table[row sep=crcr] {%
x	y\\
3	3\\
3	2\\
2.6	2.2\\
2.7	2.6\\
}--cycle;

\addplot[area legend, line width=1.0pt, draw=white, fill=white!85!black, forget plot]
table[row sep=crcr] {%
x	y\\
2.6	2.2\\
3	2\\
2	2\\
2.4	2.2\\
}--cycle;

\addplot[area legend, line width=1.0pt, draw=white, fill=white!85!black, forget plot]
table[row sep=crcr] {%
x	y\\
2	3\\
2	2\\
2.4	2.2\\
2.3	2.6\\
}--cycle;

\addplot[area legend, line width=1.0pt, draw=white, fill=white!85!black, forget plot]
table[row sep=crcr] {%
x	y\\
2.3	2.6\\
2.7	2.6\\
2.6	2.2\\
2.4	2.2\\
}--cycle;

\addplot[area legend, line width=1.0pt, draw=white, fill=mycolor37, forget plot]
table[row sep=crcr] {%
x	y\\
3	3\\
4	3\\
3.7	2.6\\
3.3	2.6\\
}--cycle;
\node[centered, align=center, inner sep=0, font=\bfseries]
at (axis cs:3.5,2.75) {6.20};

\addplot[area legend, line width=1.0pt, draw=white, fill=mycolor35, forget plot]
table[row sep=crcr] {%
x	y\\
4	3\\
4	2\\
3.6	2.2\\
3.7	2.6\\
}--cycle;
\node[centered, align=center, inner sep=0, font=\bfseries]
at (axis cs:3.825,2.45) {5.56};

\addplot[area legend, line width=1.0pt, draw=white, fill=mycolor38, forget plot]
table[row sep=crcr] {%
x	y\\
3.6	2.2\\
4	2\\
3	2\\
3.4	2.2\\
}--cycle;
\node[centered, align=center, inner sep=0, font=\bfseries]
at (axis cs:3.5,2.125) {6.81};

\addplot[area legend, line width=1.0pt, draw=white, fill=mycolor39, forget plot]
table[row sep=crcr] {%
x	y\\
3	3\\
3	2\\
3.4	2.2\\
3.3	2.6\\
}--cycle;
\node[centered, align=center, inner sep=0, font=\bfseries]
at (axis cs:3.175,2.45) {4.06};

\addplot[area legend, line width=1.0pt, draw=white, fill=mycolor36, forget plot]
table[row sep=crcr] {%
x	y\\
3.3	2.6\\
3.7	2.6\\
3.6	2.2\\
3.4	2.2\\
}--cycle;
\node[centered, align=center, inner sep=0, font=\bfseries]
at (axis cs:3.5,2.4) {1.57};
\end{axis}
\end{tikzpicture}%